\documentstyle[11pt,paspconf]{article}

\def\ltsima{$\; \buildrel < \over \sim \;$}
\def\simlt{\lower.5ex\hbox{\ltsima}} 
\def\gtsima{$\; \buildrel > \over \sim \;$}
\def\simgt{\lower.5ex\hbox{\gtsima}} 

\begin{document} 

\input{psfig.tex}

\title{Blazars: Clues to jet physics} 

\author{Rita M. Sambruna}
\affil{Penn State University, 525 Davey Lab, University Park,
PA 16802} 

\begin{abstract} 

Being dominated by non-thermal emission from aligned relativistic
jets, blazars allow us to elucidate the physics of extragalactic jets,
and, ultimately, how energy is extracted from the central black
hole. Crucial information about jet structure is provided by the
spectral energy distributions from radio to $\gamma$-rays, their
trends with luminosity, and correlated multifrequency
variability. Since blazar jets have broad implications for all
radio-loud (and possibly radio-quiet) AGNs, we also need to understand
their circumnuclear structure, especially the details of the physical
and dynamical conditions of the highly ionized gas on sub-pc scales,
which could be directly related to jet formation and radiative power.
Eventually, the bulk of information provided by blazars will help us
clarify the origin of the radio-loud/radio-quiet AGN dichotomy, one of
the most outstanding open issues of extragalactic astrophysics.

\end{abstract} 


\section{The importance of blazars for all AGNs} 

Blazars (including BL Lacertae objects and quasar-like blazars) are
among the most violent manifestations of activity in galaxies.  Their
defining properties include large luminosities emitted on short
timescales, high and variable polarization degrees in optical and IR,
smooth continuum emission from radio to $\gamma$-ray energies. These
properties are best explained as non-thermal emission from a
relativistic jet oriented close to the line of sight (Blandford \&
Rees 1978). Compelling evidence for relativistic beaming of the
emitted radiation is provided by the strong and rapidly variable
$\gamma$-ray emission observed in many blazars in recent years
(Hartman 1998).

Today, the relevance of blazars for all AGNs is increasingly
recognized as it becomes evident that relativistic jets are a feature
of accretion in all radio-loud AGNs (Urry \& Padovani 1995), and
possibly also in their radio-quiet counterparts (e.g., Falcke et al.
1996), while superluminal ejecta are also present on Galactic scales
(Mirabel \& Rodriguez 1998).  What originally made blazars appear
exotic and of little appeal to the astronomical community, now turns
out to be their major strength.  We need to understand how jets are
formed and work if we want to understand how energy is extracted from
the central black hole in AGNs (Krolik, these proc.).  Blazars provide
us with a direct probe of this fundamental issue.

\begin{figure}
\noindent {\psfig{figure=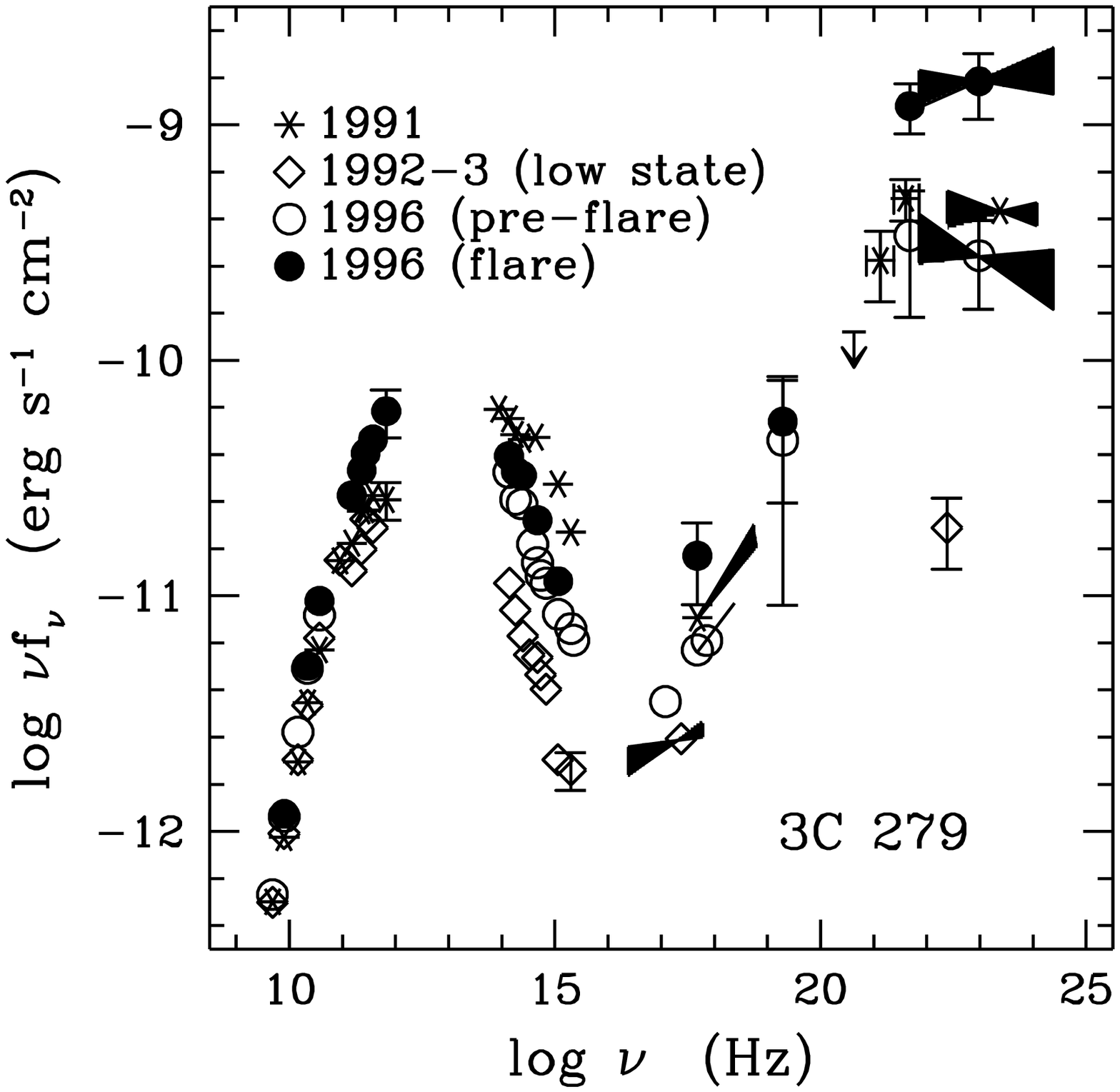,height=2.9in}}{\psfig{figure=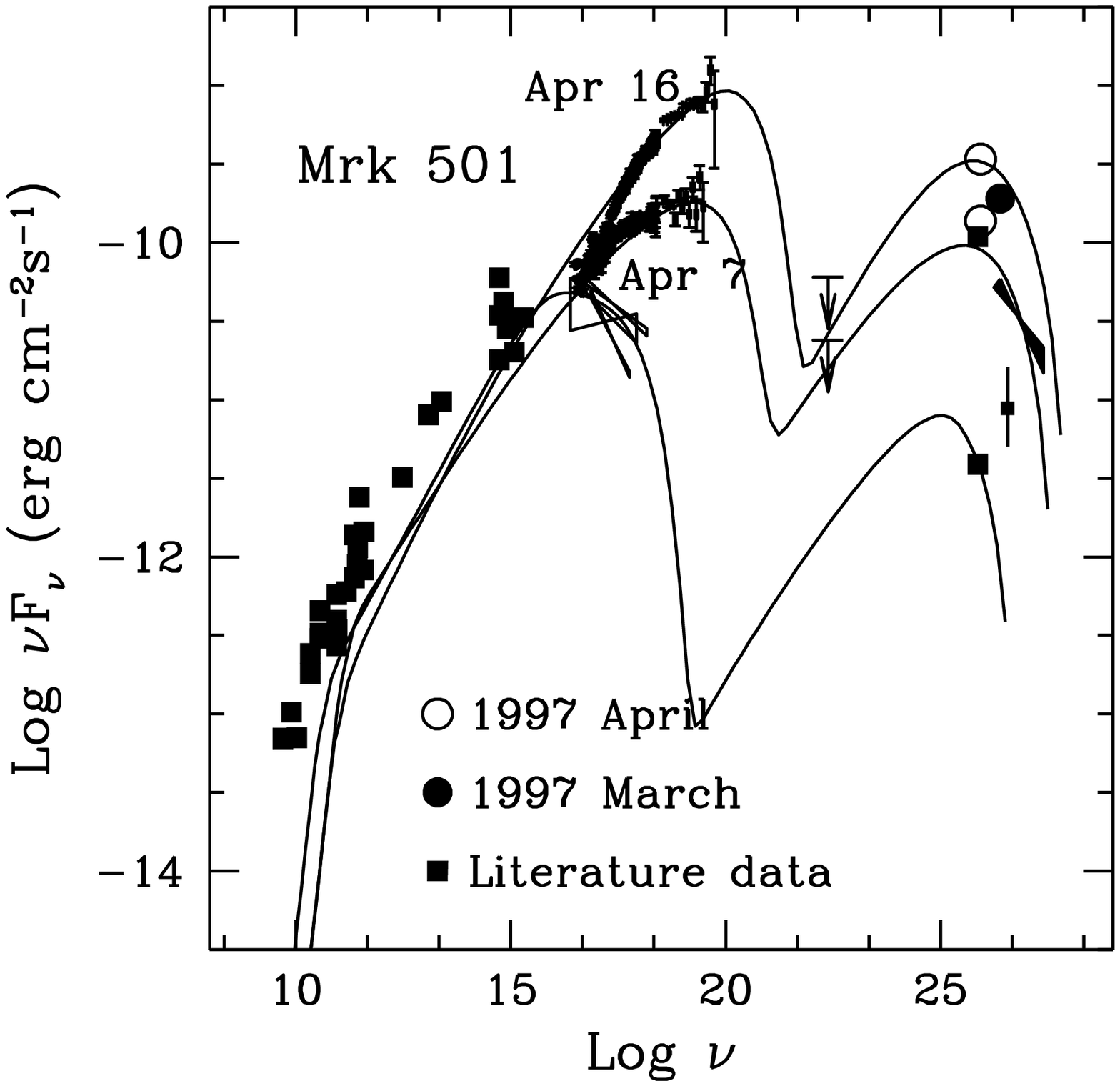,height=2.7in}}
\caption{Spectral energy distributions and variability of the red
blazar 3C279 {\it (left)} and blue blazar Mrk 501 {\it (right)}. The
continua are characterized by two broad humps peaking at different
wavelengths in the two cases, respectively. Variability is more
pronounced above the peaks at both low and high energies (data from
Wehrle et al. 1998 and Pian et al. 1998).} \label{sed}
\end{figure} 


\section{Spectral energy distributions: The blazar family}

First clues about jet physics are provided by blazars' spectral energy
distributions (SEDs).  These typically exhibit two broad humps
(Fig.~\ref{sed}): the first one peaks at IR/optical in ``red'' blazars
and at UV/X-rays in their ``blue'' counterparts (Giommi et al. 1994),
and is due to synchrotron emission (Sambruna et al. 1996).  The second
component extends from X-rays up to GeV/TeV energies, and its origin
is less well understood. One promising explanation is inverse Compton
(IC) scattering of ambient photons, either internal (synchrotron-self
Compton, SSC; Maraschi et al. 1992) or external to the jet (External
Compton, EC; Dermer \& Schlickeiser 1993; Sikora et al. 1994;
Ghisellini \& Madau 1996).

Red and blue jets are just the extrema of a continuous
population. Deep multicolor surveys are finding blazars with
intermediate spectral shapes (Laurent-Muheleisein et al. 1998; Perlman
et al. 1998).  Spectral trends with bolometric luminosity are observed
in existing samples: from red to blue blazars the luminosity decreases
and the synchrotron and IC peak frequencies decrease
proportionately. Red sources also appear to have a larger $\gamma$-ray
dominance and more luminous broad emission lines (Sambruna et
al. 1996; Sambruna 1997; Fossati et al. 1998).


In the context of a simple homogeneous model, where a single electron
population is responsible for producing both components through
synchrotron and IC processes, the spectral trends can be interpreted
as a change of a few jet physical parameters: the higher synchrotron
peak frequencies of blue blazars call for higher magnetic fields
or/and electron energies than red sources (Sambruna et al. 1996).  It
is also possible that red blazars have larger luminosities and lower
electron energies as a result of a more efficient cooling through EC
scattering (Ghisellini et al. 1998). In this case, the paradigm would
be that the type of electron cooling depends on the density of the
external photons, with the SED continuity hinting to a large spread of
physical properties of the jet ambient medium, commensurate with the
observed line properties (Scarpa \& Falomo 1997). 

However, the luminosity trends are most likely affected by strong
biases, especially at $\gamma$-rays, where the limited sensitivity of
current GeV detectors combined with large variability could exaggerate
the $\gamma$-ray dominance in the most distant red blazars (and bias
toward EC-dominated sources, where the $\gamma$-rays are more
beamed). In blue blazars Klein-Nishina cutoffs kick in at TeV energies
and the ratio of the synchrotron to IC peak is only approximately
constant. Finally, recent deep surveys show the existence of an
unexpected population of blue blazars with strong emission lines
(Perlman et al. 1998; see also Sambruna 1997). While the new spectral
trends will be tested more extensively in the future, a clear message
emerges from the current data: {\it blazars exhibit a rich variety of
spectral behaviors, which point to a diversity of jet physical
conditions} rather than being due to beaming/orientation effects only
(Sambruna et al. 1996; Georganopoulos \& Marscher 1998; Kubo et
al. 1998).

\section{Correlated multiwavelength variability: Looking inside 
the jet}

One crucial still unanswered question is the origin of the GeV and TeV
emission. An alternative to the SSC and EC scenarios (\S~2) is
provided by the proton-induced cascades models (PIC; Protheroe \&
Biermann 1997; Mannheim \& Bierman 1992), where the $\gamma$-rays are
direct synchrotron emission from ultra-relativistic electrons while
the X-rays are emitted by less energetic electrons originating from
pair cascades initiated by protons. The viability of hadronic models,
at least in red blazars, may be supported by recent observations of
circular radio polarization (Wardle et al. 1998). Clearly,
distinguishing between the IC and PIC scenarios is of central
importance to understand the jet composition and how energy is
transported away from the central black hole. 

Correlated multifrequency variability provides a way to this end,
since the different models make distinct predictions for the relative
flare amplitudes and for the time lags.  In a one-zone homogeneous
approximation, the synchrotron plus IC models predict: 1) strongly
correlated variability of the fluxes at the low- and high-energy peaks
with no time lags, since the same electron population is responsible
for emitting both spectral components; 2) simple and yet precise
relationships for the relative amplitudes of the synchrotron and IC
flares depending (for a fixed beaming factor $\delta$) on the change
of electrons and/or seed photons (Ghisellini \& Maraschi 1996); 3)
accurate shapes of the synchrotron and IC flares depending on a few
source typical timescales (Chiaberge \& Ghisellini 1998) and
parameters (Romanova \& Lovelace 1997; B\"ottcher \& Dermer 1998); 5)
lags between soft and hard X-rays, reflecting the electron radiative
time (t$_r \propto E^{-1/2}B^{-3/2}\delta^{-1/2}$), from which the
local magnetic field $B$ can be derived (modulo $\delta$; Takahashi et
al. 1996); 6) defined time-dependent spectra, according to how fast
electrons are accelerated compared to t$_r$ (Kirk et al. 1998).

In contrast, PIC models allow more freedom: lags of either sign
between long and short wavelengths are possible, as well as more
arbitrary relations in the flare relative amplitudes, because the
relative energy deposited in electrons and protons is a free
parameter, and so the details of the ensuing pair production are less
constrained.


The target selection for multifrequency campaigns is strongly
constrained by the $\gamma$-ray band, where only a handful red blazars
are bright enough to be monitored at GeV energies. At TeV even fewer
sources are detected, while TeV photons from distant blazars may be
difficult to observe because of their interaction with the IR
background (Salomon \& Stecker 1994).  Despite the intrinsic
difficulty of such campaigns, now severely undermined by the death of
the EGRET CGRO detector, a few blazars were monitored well enough to
allow comparison of models to the data.

\noindent{\bf Results for red blazars.} The multi-epoch SEDs of 3C279,
one of the most luminous GeV sources in the sky (Fig.~\ref{sed}),
shows indeed correlated variability at IR/optical and GeV, supporting
IC models. The amplitude of the GeV variation goes quadratically with
the optical flux in early campaigns, indicating a variation of the
electron density as the cause of the flare in the SSC model (Maraschi
et al. 1994) or a change of beaming factor in the EC model.  During
the 1996 campaign, however, the $\gamma$-rays varied more than
quadratically than the optical/IR flux, and exactly (within 1 day)
simultaneously to the X-rays (Wehrle et al. 1998), with the rapid
decay time of the GeV flare favoring EC scenarios (Ghisellini \& Madau
1996).

A new red candidate for multifrequency campaigns is BL Lac itself,
which underwent a strong GeV and optical outburst in 1997 July (Bloom
et al. 1997). The $\gamma$-ray light curve shows a strong flare
possibly anticipating the optical flare by $\sim$ 0.5 days; however,
the poor $\gamma$-ray sampling prevents any firm
conclusion. Interestingly, a broad H$\alpha$ emission line was
detected in BL Lac (Vermeulen et al. 1995), which increased in
luminosity at the time of the optical/GeV outburst, suggesting that EC
models are responsible for producing the flux above a few MeV. This is
indeed supported by a detailed modeling of the SED (Sambruna et
al. 1999). 


\noindent{\bf Results for blue blazars.} Mrk 501, one of only four
extragalactic TeV sources (all blazars), flared dramatically at TeV
and X-rays in 1997 April (Aharonian et al. 1997; Catanese et
al. 1997); TeV and X-ray light curves track each other well, with no
lags larger than 1 day, confirming that the TeV photons are produced
by the same electrons responsible for the X-rays. A similar behavior
was observed also during our recently concluded campaign in 1998
June. As shown in Fig.~\ref{sed}, during the 1997 TeV flare the
synchrotron peak shifted forward by 2 orders of magnitude in
correspondence to a dramatic flattening of the X-ray continuum,
implying large acceleration events (Pian et al. 1998). Mrk 501 is an
eloquent example of the dramatic powers that can be reached in
blazars. 

An ideal candidate for studying the energy-dependence of the
synchrotron flares is PKS 2155-304, one of the brightest and most
rapidly variable BL Lacs at optical through X-ray wavelengths,
recently detected in TeV (Chadwick et al. 1998).  Previous campaigns
(Edelson et al. 1995; Urry et al. 1997) detected correlated
variability at optical through X-ray wavelengths, and established that
the shorter wavelengths generally vary first, with lags of the order
of a few hours to a few days. The measured 1 hour delay between soft
and hard X-rays in 1994 May yields $B \sim 0.1$ Gauss (Urry et
al. 1997), similar to Mrk 421 (Takahashi et al. 1996).

\begin{figure}
\noindent {\psfig{figure=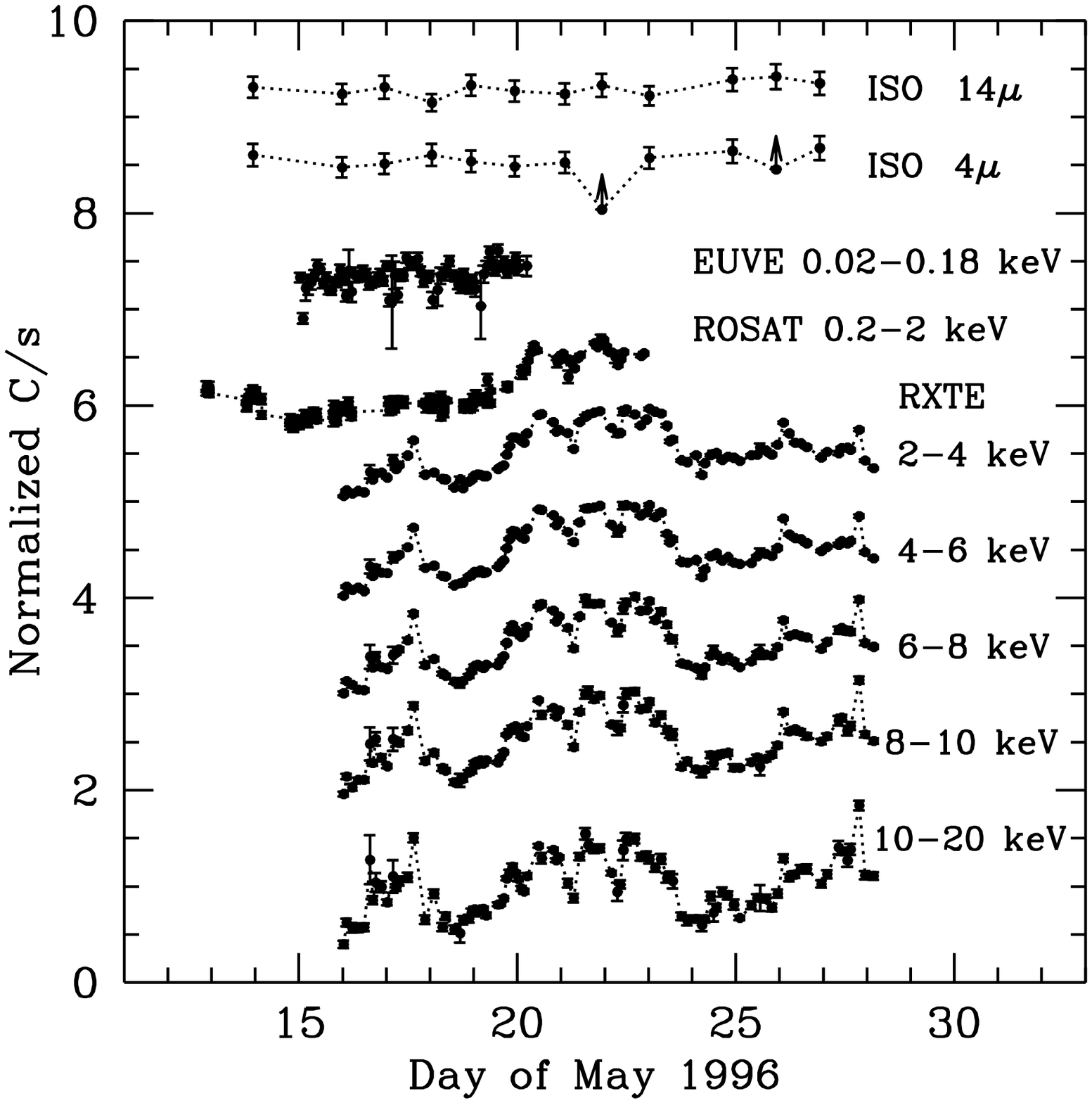,height=2.9in}}{\psfig{figure=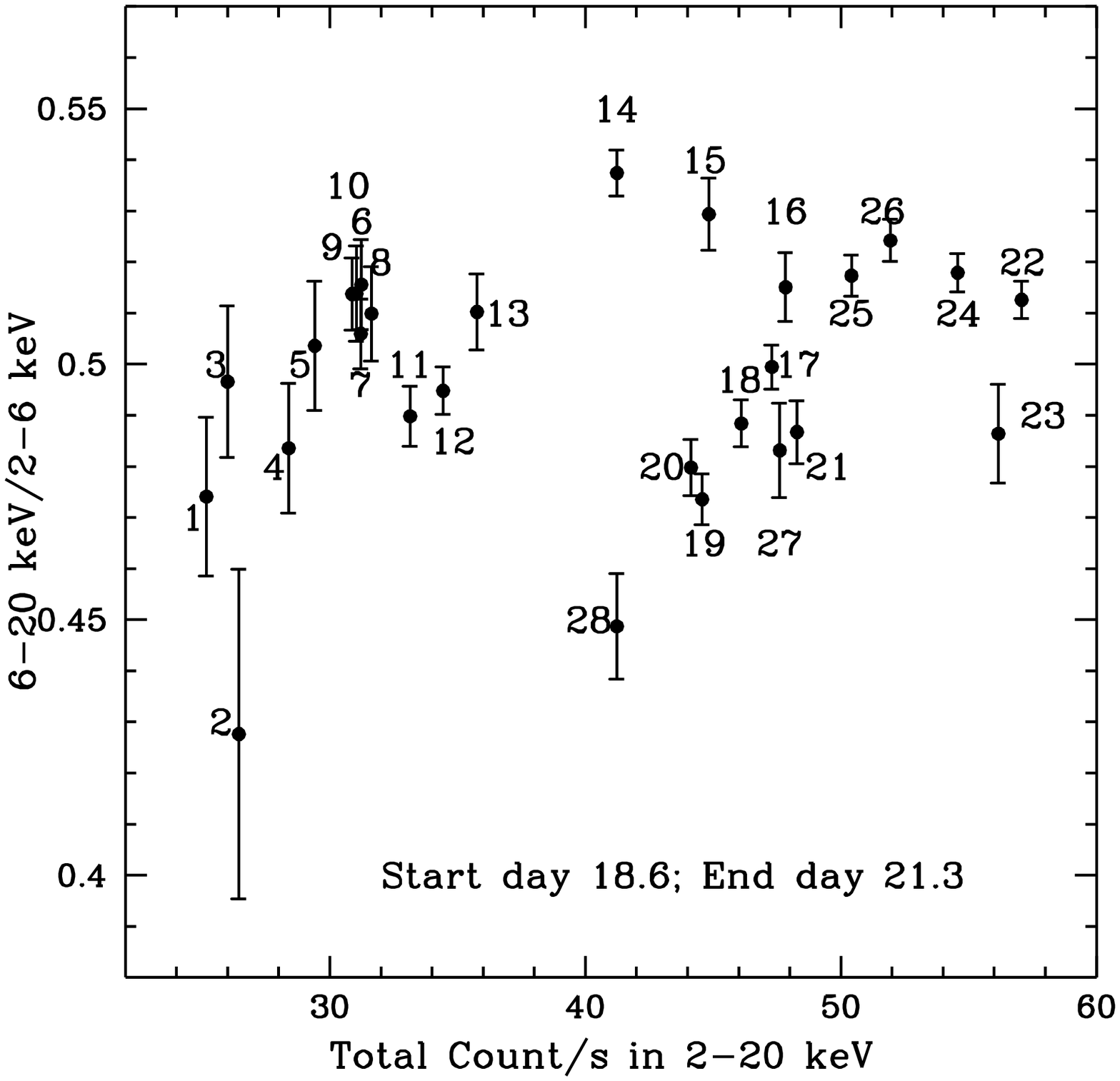,height=2.9in}}
\caption{Multifrequency variability of PKS 2155--304 in 1996 May. {\it
Left:} IR to X-ray light curves, normalized to their mean fluxes and
arbitrarily offset (data from papers in prep. by Bertone, Brinkmann,
Marshall, Sambruna). No lags are detected larger than the RXTE binning
time (90 min); the flares have larger amplitudes and more symmetric
shapes at increasing energies. {\it Right:} An example of X-ray
hardness ratios vs. intensity, numbered chronologically. The spectral
changes follow a general clock-wise cycle, with anti-clock-wise
smaller loops. This is a diagnostic of the acceleration processes in
the emission region.} 
\label{2155}
\end{figure}

The multifrequency experiment for PKS 2155--304 was repeated in 1996
May (Fig.~\ref{2155}), with better sampling in X-rays but worse at
longer wavelengths. Complex variability was observed, with several
short flares superposed to a longer trend. The flares become more
symmetric at decreasing wavelengths, roughly consistent with a
homogeneous synchrotron scenario where the radiative time at high
energies is fast, and the light travel time limits both the rise and
decay times (Chiaberge \& Ghisellini 1998). No lags larger than $\sim$
1.5 hours were observed. The X-ray hardness ratios (Fig.~\ref{2155})
follow a general clock-wise loop with intensity, indicating that
acceleration is fast and the spectrum is controlled by radiative
cooling, with superposed events of anti-clockwise cycles, when cooling
and acceleration are in equilibrium (Kirk et al. 1998). 

\noindent{\bf What did we learn?} Current multifrequency data support
models which predict correlated variability at the low- and
high-energy peaks in both blue and red blazars. While it proved
difficult to distinguish among models based on the relative flare
amplitudes, due to the large number of free parameters involved,
detection of time lags and flare shape promises to be more efficient
to this end, but is limited at present by sampling constraints,
especially at $\gamma$-rays. Spectral variability at X-rays is a
powerful diagnostic of the acceleration events in the jets, and can be
studied well with RXTE and SAX.

\section{Blazar environment: Where jets form} 

The study of blazar environment has recently received much attention,
as the importance of the external medium for jet formation and
deceleration (De Young 1993; Bicknell 1995) and for jet radiative
power (Dermer \& Chiang 1998) is increasingly recognized. IR and
optical studies from the ground (see Heidt 1998 for a review) and with
HST (Urry et al. 1999 and references therein) show that BL Lacs
generally lie in poor clusters (Abell richness 0--1), and that their
host galaxies are smooth, luminous ellipticals, with no differences
between red and blue sources. The host galaxies of BL Lacs are very
similar to those of their putative parent populations, Fanaroff-Riley
I galaxies, of matching extended radio powers (Urry et al. 1999b, in
prep.), supporting unified models for radio-loud AGNs.  Substantial
progress in the study of blazar environs will be achieved in the near
future with AXAF\footnote[1]{The Advanced X-ray Astrophysical Facility
will carry, among its various detectors, a set of CCD imagers with
spectroscopic capabilities, with a much narrower PSF (FWHM \simlt 1
arcsec) and lower intrinsic background than ROSAT.} observations in
X-rays, which will allow us to measure directly the medium properties
(temperature, density, abundances) and probe in depth the
gravitational potential.

\noindent{\bf High-energy absorption in BL Lacs.} A powerful probe of
the inner (pc/sub-pc scales) jet environs is provided by
high-resolution observations of a few bright BL Lacs at X-ray and EUV
wavelengths. Broad absorption features were detected in PKS 2155--304
and Mrk 421, the only two blazars observed at grating resolutions in
soft X-rays (Canizares \& Kruper 1984) and EUV (K\"onigl et al. 1995;
Kartje et al. 1997), suggesting absorption in highly ionized gas
outflowing with mildly relativistic velocities ($v$ \simlt $0.1c$).
Ionized absorption was also detected in BBXRT and ASCA observations of
a few more BL Lacs (Sambruna et al. 1997; Sambruna \& Mushotzky 1998),
confirming earlier claims (Madejski et al. 1992).

The origin of the absorbing medium is still a speculation.  One
possibility is the disk-driven hydromagnetic wind model (K\"onigl \&
Kartje 1994), based on the mechanism originally proposed by Blandford
\& Payne (1982) to explain the origin of relativistic jets (Lovelace,
these proc.): clouds of gas are lifted from the accretion disk surface
by the magnetic field lines until they intercept the beamed jet
radiation, giving rise to absorption when crossing the line of
sight. Alternatively, absorption could originate in matter swept up by
the jet outer edge (Krolik et al. 1985), perhaps the same decelerating
medium required by unification schemes (Bicknell 1995). At the current
low spectral resolution and sensitivity there is no way to
discriminate among the possible models; however, this will be trivial
with the gratings onboard the future AXAF and X-ray Multi-Mirror
missions, which will allow us to measure precisely the absorber
properties and trace a detailed portrait of the inner jet environs.

\section{Future work} 

While recent years have seen gigantic progress in our understanding of
the blazar phenomenon, much more work remains to be done.

We need to test more extensively the luminosity/distance spectral
trends, in particular at $\gamma$-rays, with larger complete samples,
fully addressing the various selection effects.  More TeV
observations, where we currently have only 4 detected sources, are
badly needed, as well as more precise determinations of the TeV
spectra in order to quantify better the position of the IC peak in
blue blazars.

We need to intensify our efforts in multiwavelengths monitorings of
blue and (more) red blazars, in particular to measure more precisely
the time lags down to the smallest accessible scales, and the flare
shapes both at low and high energies.  With the death of EGRET on
CGRO, the short-term future of multifrequency campaigns relies only on
TeV blazars.

Blazar environment studies at IR, optical, and X-ray wavelengths with
current and future high-resolution satellites (HST, NGST, AXAF, XMM)
will address the AGN/host galaxy connection, in particular the role of
the ambient medium for jet formation and collimation, and for fueling
the central black hole. 

With the advent of new technical resources of improved sensitivity and
resolution in a large wavelength range, the future holds great promise
for blazars. Eventually, the bulk of information provided by blazars
will help us shed light on the origin of jets and, ultimately, of the
radio-loud/radio-quiet AGN dichotomy, one of the most outstanding open
problems of extragalactic astrophysics.

\acknowledgments 

This work was supported by NASA grants NAS--38252, NAG--3313,
NAG5--7121, and by an IAU travel grant. I thank Meg Urry for
interesting discussions, and all my colleagues at the conference for a
memorable meeting.

\begin{question}{Dave Sanders}
You showed a nice plot [extended radio power vs. host magnitude]
indicating that most blazars are in FRIs. Does this have implications
for understanding possible evolution of FRIs and FRIIs?
\end{question}
\begin{answer}{Rita Sambruna}
The data I showed (from Urry et al. 1999b, in prep.) are most for BL
Lacs at weaker intrinsic powers, which are all nearby blue
blazars. Before answering your question we need to know where red
blazars, especially those with strong lines (the classical FSRQs),
will fall on the diagram, likely with FRIIs.
\end{answer} 

\begin{question}{Vahe Petrosian}
Does the variability of 3C279 agree with the trend between the
bolometric luminosity and peak frequency? Does the observed lag in PKS
2155--304 agree with both the homogeneous and inhomogeneous models?
\end{question}
\begin{answer}{Rita Sambruna}
3C279 is a perfect example of the EGRET selection effects for the
trends. It does not provide compelling evidence that they are wrong;
most depends on the duty cycle of the $\gamma$-ray high state.  The
time lag in PKS 2155--304 in 1994 can be explained in the homogeneous
model in terms of radiative losses (high-energy particles cool off
first) or some type of energy stratification, behind a shock or/and in
a inhomogeneous model (Urry et al. 1997).
\end{answer}

\begin{question}{Ian Robson}
Are you sure the 1997 optical/UV/X-ray component in Mrk 501 is really
an extension of the previous synchrotron component rather than
something new? 
\end{question}
\begin{answer}{Rita Sambruna}
The observed spectral X-ray variability fits with what we have always
observed in this and other blue blazars: flatter continuum with
increasing flux, which is naturally explained in the context of
synchrotron emission in terms of injection of fresh particles or a
strong reacceleration event. 
\end{answer}

\begin{question}{Dan Weedman}
Do blazars exhibit any evidence for accretion disk activity, which
could trigger the jet? 
\end{question}
\begin{answer}{Rita Sambruna}
The disk emission in blazars is very difficult to observe because the
jet enhanced emission overshines it. In the few cases where we do see
a blue bump (e.g., 3C345), the latter is only a small fraction of the
total UV flux.
\end{answer} 

\begin{question}{Halton Arp}
Red blazars should be the major contributors of cosmic rays. 
\end{question}
\begin{answer}{Rita Sambruna}
They are certainly strong high-energy emitters, and major contributors
to the GeV background. They could also be TeV emitters, but are too
distant to be detected. 
\end{answer}


\begin{references}
\reference Aharonian, F. et al. 1997, \astap, 327, L5
\reference Bicknell, G. V., 1995, \apjsupp, 101, 29 
\reference Blandford, R.D. \& Payne, D.G. 1982, \mnras, 199, 883 
\reference Blandford, R.D. \& Rees, M.J. 1978, in Pittsburgh
Conference on BL Lac Objects, A.M.Wolfe, Univ. Pittsburgh Press, 1978,
p.328 
\reference Bloom, S. D. et al. 1997, \apjlett, 490, L145
\reference B\"ottcher, M. \& Dermer, C.D. 1998, \apjlett, 501, L51 
\reference Canizares, C.R. \& Kruper, J. 1984, \apjlett, 278, L99 
\reference Catanese, M. et al. 1997, \apjlett, 487, L143 
\reference Chadwick, P.M. et al. 1998, \apj, in press, astro-ph/9810209 
\reference Chiaberge, M. \& Ghisellini, G. 1998, \mnras, subm., 
astro-ph/9810263 
\reference De Young, D. S. 1993, \apjlett, 405, L13 
\reference Dermer, C.D. \& Chiang, J. 1998, astro-ph/9810222 
\reference Dermer, C.D. \& Schlickeiser, R. 1993, \apj, 416, 458
\reference Edelson, R. et al. 1995, \apj, 438, 120 
\reference Falcke, H. et al. 1996, \apjlett, 473, L13 
\reference Fossati, G. et al. 1998, \mnras, 299, 433 
\reference Georganopoulos, M. \& Marscher, A. 1998, \apj, 506, 621 
\reference Ghisellini, G. et al. 1998, \mnras, in press, astro-ph/9807317  
\reference Ghisellini, G. \& Madau, P. 1996, \mnras, 280, 67 
\reference Ghisellini, G. \& Maraschi, M. 1996, in Blazar
Continuum Variability, H.R.Miller, J.R.Webb, \& J.C.Noble, ASP
Conf. Series, Vol. 110, 436 
\reference Giommi, P., Ansari, S.G., \& Micol, A. 1995, A\&ASupp., 109, 267
\reference Hartman, R.C. 1998, in BL Lac Phenomenon,
Eds. A.Sillanpaa \& L.O.Takalo, PASP Conf. Series, in press 
\reference Heidt, J. 1998, in BL Lac Phenomenon, Eds. A.Sillanpaa \& 
L.O.Takalo, PASP Conf. Series, in press 
\reference Kartje, J.F. et al. 1997, \apj, 474, 630 
\reference Kirk, J.G., Rieger, F.M., \& Mastichiadis, A. 1998, \astap,
333, 452 
\reference K\"onigl, A. et al. 1995, \apj, 446, 598 
\reference K\"onigl, A. \& Kartje, J.F. 1994, \apj, 434, 446
\reference Krolik, J.H. et al. 1985, \apj, 295, 104 
\reference Kubo, H. et al. 1998, \apj, 504, 693 
\reference Laurent-Muehleisen, S.A. et al. 1998, \apjsupp, 118, 127
\reference Madejski, G. et al. 1991, \apj, 370, 198
\reference Mannheim, K. \& Biermann, P.L. 1992, \astap, 253, L21 
\reference Maraschi, L. et al. 1994, \apjlett, 435, L91 
\reference Maraschi, L., Ghisellini, G., \& Celotti, A. 1992, \apj, 397, L5 
\reference Mirabel, I.F. \& Rodriguez, L.F. 1998, Nature, 392, 673 
\reference Perlman, E.S. et al. 1998, \aj, 115, 1253 
\reference Pian, E. et al. 1998, \apjlett, 492, L17 
\reference Protheroe, R.J. \& Bierman, P.L. 1997, APh, 6, 293
\reference Romanova, M.M. \& Lovelace, R.V.E. 1997, \apj, 475, 97 
\reference Salomon, H.M. \& Stecker, F.W. 1994, \apjlett, 430, L21 
\reference Sambruna, R.M. et al. 1999, \apj, in press, astro-ph/9810319 
\reference Sambruna, R.M. \& Mushotzky, R.F. 1998, \apj, 502, 630
\reference Sambruna, R.M. 1997, \apj, 474, 639 
\reference Sambruna, R.M. et al. 1997, \apj, 483, 774 
\reference Sambruna, R.M., Maraschi, L., \& Urry, C.M. 1996, \apj, 463, 444
\reference Scarpa, R. \& Falomo, R. 1997, \astap, 325, 109   
\reference Sikora, M., Begelman, M.C., \& Rees, M.J. 1994, \apj, 421, 153 
\reference Takahashi, T. et al. 1996, \apjlett, 470, L89
\reference Urry, C.M. et al. 1999, \apj, in press, astro-ph/9809030 
\reference Urry, C.M. et al. 1997, \apj, 486, 799 
\reference Urry, C.M. \& Padovani, P. 1995, \pasp, 107, 803 
\reference Vermeulen, R.C. et al. 1995, \apj, 452, L5 
\reference Wardle, J.F.C. et al. 1998, Nature, 395, 457 
\reference Wehrle, A. E. et al. 1998, \apj, 497, 178 
\end{references}
\end{document}